\documentclass{article}

\usepackage{arxiv}
\usepackage[numbers]{natbib}

\usepackage{amsmath, enumitem}
\usepackage{upgreek}
\usepackage{amsfonts, bm}
\usepackage{upgreek}
\usepackage{graphicx}
\graphicspath{{Figures/}} % Location of the graphics files
\usepackage{setspace}
\usepackage{multirow}
\usepackage[T1]{fontenc}

\DeclareMathOperator*{\argmax}{arg\,max}

\title{Interrogating probabilistic inversion results for subsurface structural information}

\author{
	Xin Zhang \\
	School of Geosciences \\
	University of Edinburgh\\
	Edinburgh, United Kingdom \\
	\texttt{x.zhang2@ed.ac.uk} \\
	\And
	Andrew Curtis \\
	School of Geosciences \\
	University of Edinburgh\\
	Edinburgh, Unite Kingdom \\
	\texttt{andrew.curtis@ed.ac.uk} \\
}
\begin{document}
	\maketitle

\begin{abstract}
	The goal of a scientific investigation is to find answers to specific questions. In geosciences this is typically achieved by solving an inference or inverse problem and interpreting the solution. However, the answer obtained is often biased because the solution to an inverse problem is nonunique and human interpretation is a biased process. Interrogation theory provides a systematic way to find optimal answers by considering their full uncertainty estimates, and by designing an objective function that defines desirable qualities in the answer. In this study we demonstrate interrogation theory by quantifying the size of a particular subsurface structure. The results show that interrogation theory provides an accurate estimate of the true answer, which cannot be obtained by direct, subjective interpretation of the solution mean and standard deviation. This demonstrates the value of interrogation theory. It also shows that fully nonlinear uncertainty assessments may be critical in order to address real-world scientific problems, which goes some way towards justifying their computational expense.
\end{abstract}

\section{Introduction}
Geoscientists often wish to find answers to specific scientific questions: How large is a subsurface body? How deeply does lithosphere subduct? How likely is this volcano to erupt? What method provides the most accurate results? To answer such questions, background research is conducted to reveal existing information, and experiments are designed and performed to acquire new data. The answer to the question is then estimated by interpreting this information. 

Answering a question therefore requires that we obtain useful information relevant to the answer from both existing information (often called the \textit{prior} information) and new data. This usually involves solving an inference or inverse problem \citep{tarantola2005inverse}. For example, to answer questions about the Earth's interior scientists often build subsurface tomographic models from data observed at the surface: seismic velocity structures are obtained from seismic data, or resistivity structures may be constructed from electromagnetic data. This involves solving an inverse problem to estimate a subsurface model or family of models that are consistent with the data, and an answer to the question may be interpreted from the solution. We address such a case herein.

Due to the nonlinear physical relationship between model parameters and data, insufficient data coverage and noise in the data, the inverse problem almost always has nonunique solutions as many sets of model parameter values fit the data to within their measurement uncertainties. It is therefore important to characterize the uncertainty of a solution such that the final answer can take into account the range of possible models.

The tomographic inverse problem is often solved using either standard optimization or Bayesian inference. In optimization one seeks a solution which minimizes a misfit function between the observed data and the data predicted from the parameter values of a model \citep{tarantola2005inverse, aster2018parameter}. However, since the method only finds one single set of parameter values, it is difficult to characterize the uncertainty and hence the information value of the solution. As a result the answer obtained from that solution can be biased. Bayesian inference provides a different way to solve the inverse problem. In Bayesian inference one constructs a probability density function (pdf) that describes the uncertainty of solutions, called the \textit{posterior} pdf, by combining the prior information and the information contained in the data. Statistics or samples of that distribution are estimated in order to characterise the solution. Methods used for Bayesian inference include Monte Carlo sampling \citep{brooks2011handbook, sambridge2002monte} and variational inference \citep{blei2017variational,nawaz2018variational, zhang2020seismic}.

While many studies have been conducted to solve inverse problems, few have found answers to specific scientific questions based on the solutions. In practice those questions are typically answered by subjectively interpreting the solution of inverse problems using either the optimal solution, or using the mean of the posterior pdf together with the standard deviation structure. Since these statistics do not represent the full uncertainty in the solutions, and since human interpretation is a biased process \citep{polson2010dynamics}, the answer obtained in such a way is likely to be biased and does not take account of the full uncertainty.

To resolve this issue, \citet{arnold2018interrogation} introduced \textit{interrogation theory} which provides a systematic way to answer specific questions. With interrogation theory data are acquired in the way that best answers the question, multiple models are designed and discriminated, parameters of each model can be constrained by multiple algorithms, and optimal answers are found using decision theory.

\citet{arnold2018interrogation} is a complex paper that introduces interrogation theory in a general way. So in this study we demonstrate how to apply interrogation theory to estimate the size of a subsurface velocity structure obtained using seismic full-waveform inversion (FWI). We solve the inverse problem using Stein variational gradient descent (SVGD) to produce the full Bayesian posterior pdf. The size of the velocity structure is then estimated by interrogating the obtained posterior pdf. In the following section we first review concepts of interrogation theory. In section 3 we use the theory to estimate the size of a subsurface velocity structure obtained using FWI. The results show that the estimated size is very close to the true value, a result that cannot be obtained using only one single velocity structure. This demonstrates that the fully nonlinear estimates of uncertainty are critical for decision-making in real-world problems, thus justifying their computational expense. 

\section{Methods}
\begin{figure}
	\includegraphics[width=1.0\linewidth]{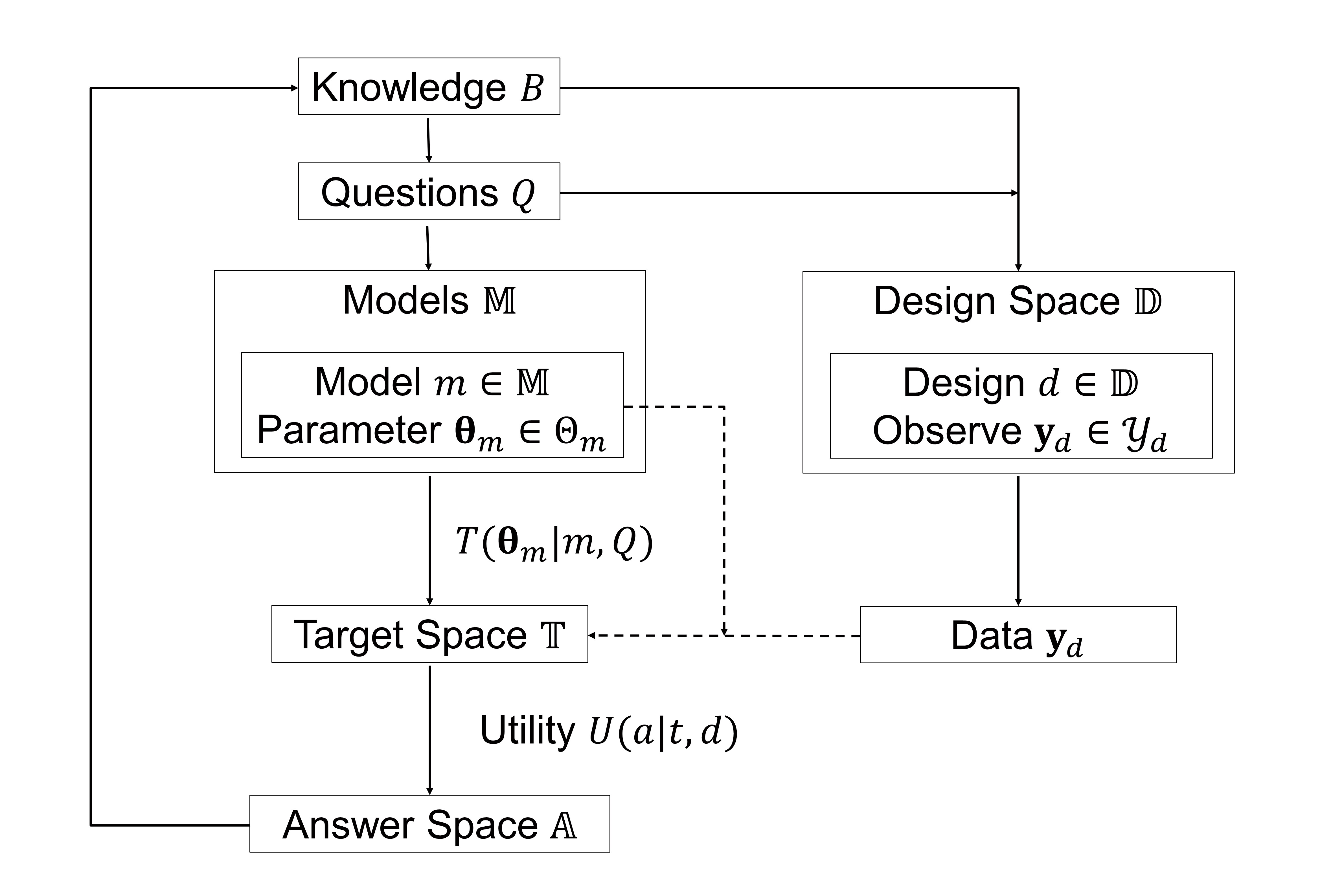}
	\caption{Algorithmic scheme of interrogation theory.}
	\label{fig:scheme}
\end{figure}
\subsection{Interrogation theory}
Interrogation theory provides a systematic way to find optimal answers to specific questions. Figure \ref{fig:scheme} shows an algorithmic scheme for the theory. An investigator has prior knowledge $B$ and wishes to answer a question $Q$. There is a set of answers $\mathbb{A}$ among which a choice needs to be made. For example, in geophysics scientists may ask questions about the depth of moho or the size of a reservoir, for which the answer space $\mathbb{A}$ will contain a set of real-numbered values. In other cases we may want to know whether a specific geology structure exists, and the answer space will contain only two values: yes or no. 

In order to answer the question, the investigator defines a set of models $\mathbb{M}$ that are relevant to the investigation. Similarly to \citet{arnold2018interrogation} here we use the term  "model" in a mathematical sense to mean a relationship between the observed data and the parameters of the model. For example, in full-waveform inversion (FWI) we use a model $m \in \mathbb{M}$ consisting of parameters $\bm \uptheta_m$ in parameter space $\Theta_m$ which represent a 3D seismic velocity structure of the Earth's interior, together with the relationship between this velocity structure and seismic waveforms. The model space $\mathbb{M}$ is therefore related to our prior knowledge $B$ and the question $Q$. In general an element or a set of elements in this space are assumed to provide a sufficiently accurate description of the state of nature relevant to answering $Q$.

To answer $Q$, an investigator needs to collect new information, which involves designing experiments and collecting data. Here we use $\mathbb{D}$ to denote the design space that contains all possible experimental designs. For each design $d \in \mathbb{D}$ there is a data space $\mathcal{Y}_{d}$ which contains all possible observations $\mathbf{y}_{d} \in \mathcal{Y}_{d}$. After an experiment a single dataset $\mathbf{y}_{d}^{obs}$ will have been collected. 

For some questions $Q$ there may be many different relevant models in $\mathbb{M}$. For example, to study subsurface Earth structure one can use seismic data to infer the seismic velocity structure, or use resistivity data to infer the resistivity structure. However no matter which model is used, it must allow us to answer the question $Q$ given the model parameters $\bm \uptheta_{m}$. Thus for each model $m$ and the question $Q$ there exists a target function $T(\bm \uptheta_{m}|m,Q)$ which maps the model parameters $\bm \uptheta_{m}$ to a target space $\mathbb{T}$ that is common for all possible models $\mathbb{M}$, and where the function $T$ summarizes only the information needed to answer question $Q$.

To find the optimal answer in the answer space $\mathbb{A}$, we define a utility function $U(a|t,d)$ which defines the benefit associated with accepting an answer $a$ given the summarized state $t=T(\bm \uptheta_{m}|m,Q)$ and the design $d$. The utility function is conditioned on $d$ so that the benefits can account for the cost of conducting the experiment with design $d$. An optimal answer is found by maximizing this utility function in the answer space $\mathbb{A}$, and several analytic results that aid this calculation are given by \citet{arnold2018interrogation}.

\subsection{Optimal answers and designs}
Given a specific experimental design $d$ and the observed data $\mathbf{y}_{d}$, the optimal answer to a question $Q$ is found by maximizing the investigator's utility function $U$. In principle this problem may be solved by maximizing a utility function $U(a|\bm \uptheta_{m},m,\mathbf{y}_{d},d)$ for answer $a \in \mathbb{A}$, parameters $\bm \uptheta_{m} \in \Theta_{m}$ embedded in model $m$, and data $\mathbf{y}_{d}$ collected under a design $d$. However, such a utility function $U(a|\bm \uptheta_{m},m,\mathbf{y}_{d},d)$ is difficult to specify when considering the spectrum of different parameters $\bm \uptheta_{m}$ of different models $m$, and different data $\mathbf{y}_{d}$ under different designs $d$. To resolve this issue, \citet{arnold2018interrogation} introduced the above target space $\mathbb{T}$ which is a common space for all possible models. The utility function can then be defined on the level of this common space, that is $U(a|t,d)$, which avoids the need to specify a utility function for every parameter value for every model in the model space.

Define $p(\bm \uptheta_{m},m|\mathbf{y}_{d},d)$ as the Bayesian posterior probability density function (pdf) for a model $m$ and its parameters $\bm \uptheta_{m}$ given the observed data $\mathbf{y}_{d}$ under the design $d$. According to Bayes theorem,
\begin{equation}
	p(\bm \uptheta_{m},m|\mathbf{y}_{d},d) = \frac{p(\mathbf{y}_{d}|\bm \uptheta_{m},m,d)p(\bm \uptheta_{m}|m)p(m)}{p(\mathbf{y}_{d}|d)}
	\label{eq:bayes}
\end{equation}
where $p(\mathbf{y}_{d}|\bm \uptheta_{m},m,d)$ is the likelihood function of observing data $\mathbf{y}_{d}$ given parameters $\bm \uptheta_{m}$, embodied in a model $m$ and under the design $d$. $p(\bm \uptheta_{m}|m)$ is the prior pdf of $\bm \uptheta_{m}$ associated with a model $m$ and $p(m)$ is the prior pdf of model $m$. $p(\mathbf{y}_{d}|d)$ is a normalization factor called the \textit{evidence}. The posterior pdf on the left of equation \ref{eq:bayes} can be obtained by Bayesian inference, for example by using Monte Carlo sampling or variational inference methods. The expected posterior utility of answers can then be constructed by integrating or summing over the set of models and over the parameter space of each model:
\begin{equation}
	U_{p}(a|\mathbf{y}_{d},d) = \sum_{m \in \mathbb{M}} \int_{\Theta_{m}} U(a|T(\bm \uptheta_{m}|m),d) p(\bm \uptheta_{m},m|\mathbf{y}_{d},d) \mathrm{d}\bm \uptheta_{m}
	\label{eq:expected_utility}
\end{equation} 
The optimal answer $a^{*}(\mathbf{y}_{d},d)$ is obtained by maximizing the expected utility function:
\begin{equation}
	a^{*}(\mathbf{y}_{d},d) = \argmax_{a \in \mathbb{A}} U_{p}(a|\mathbf{y}_{d},d)
	\label{eq:optimal_answer}
\end{equation}
The corresponding maximized utility function is $U^{*}(\mathbf{y}_{d},d) = U_{p}(a^{*}|\mathbf{y}_{d},d)$.

If the investigator wishes to find the best experimental design, the solution can be obtained similarly by maximizing the utility function in the design space. First, the expected utility for each design $d$ can be obtained by integrating over the data space:
\begin{equation}
	U^{*}(d) = \int_{\mathcal{Y}_{d}} U^{*}(\mathbf{y}_{d},d) p(\mathbf{y}_{d}|d) \mathrm{d} \mathbf{y}_{d}
	\label{eq:expected_utility_design}
\end{equation}
where $p(\mathbf{y}_{d}|d)$ represents the probability of observing data $\mathbf{y}_{d}$ under the design $d$. This distribution corresponds to the evidence in Bayes theorem, and is obtained by integrating over the model space and over the parameter space of each model:
\begin{equation}
	p(\mathbf{y}_{d}|d) = \sum_{m \in \mathbb{M}} \int_{\Theta_{m}} p(\mathbf{y}_{d}|\bm \uptheta_{m},m,d)p(\bm \uptheta_{m}|m)p(m) \mathrm{d}\bm \uptheta_{m}
	\label{eq:evidence}
\end{equation}
The best experimental design is then obtained by maximizing the utility in equation \ref{eq:expected_utility_design}:
\begin{equation}
	d^{*} = \argmax_{d \in \mathbb{D}} U^{*}(d)
	\label{eq:optimal_design}
\end{equation}

We now define the utility function $U(a|t,d)$. In the simplest case we can assume the target space and the answer space are identical; that is, our question is to estimate the summarized state $t=T(\bm \uptheta_{m}|m)$. In this case we might define the utility function as:
\begin{equation}
	U(a|t,d) = U(a|t) = - (t-a)^{2}
	\label{eq:utility}
\end{equation}
where in the first equality we have neglected the cost of conducting the experiment with design $d$. This utility is maximized when the answer $a$ equals to the true state $t$.  \citet{arnold2018interrogation} showed that in this case the optimal answer is:
\begin{equation}
	a^{*}(\mathbf{y}_{d},d) = E[T|\mathbf{y}_{d},d] = \sum_{m \in \mathbb{M}} \int_{\Theta_{m}} T(\bm \uptheta_{m}|m) p(\bm \uptheta_{m},m|\mathbf{y}_{d},d) \mathrm{d}\bm \uptheta_{m}
	\label{eq:optimal_answer_simple}
\end{equation}
which states that the optimal answer is the posterior mean of the summarized state $T$. Using different utility functions in equation \ref{eq:utility}, or answering different types of question (e.g., categorical questions), results in different forms for the optimal answer in equation \ref{eq:optimal_answer_simple}.

\section{Results}

\begin{figure}
	\includegraphics[width=1.\linewidth]{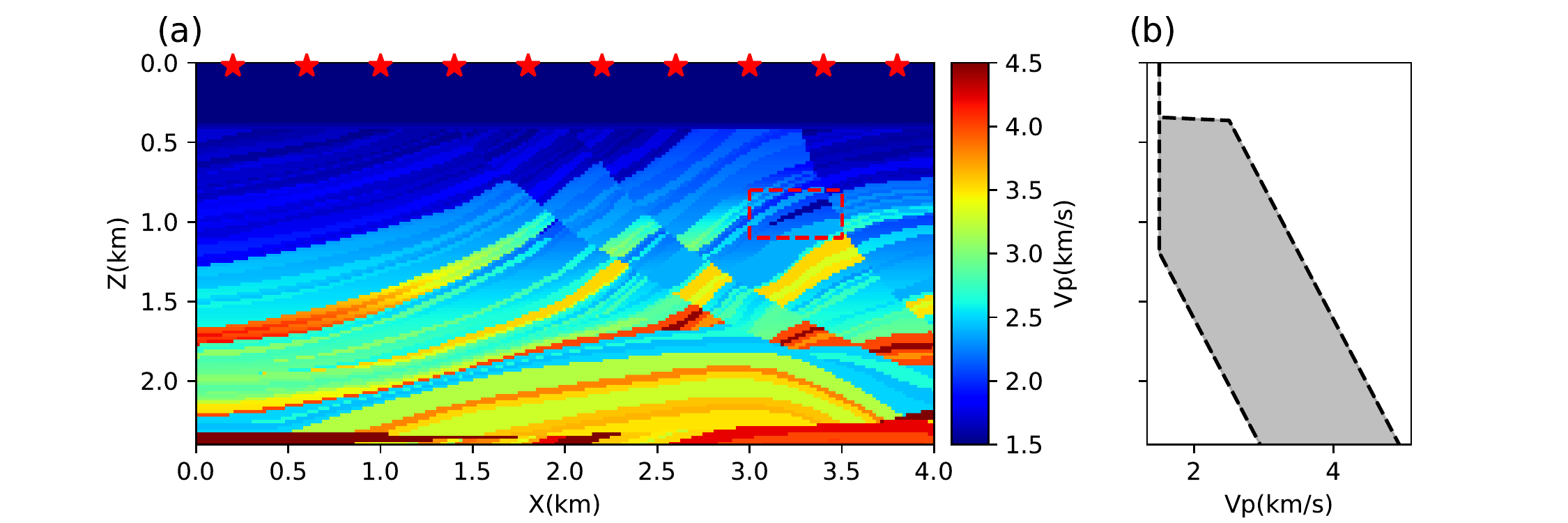}
	\caption{\textbf{(a)} The true velocity model. Red stars denote locations of 10 sources. The 200 receivers are equally spaced along the seabed at 0.36 km depth. \textbf{(b)} The prior distribution of seismic velocity, which is chosen to be a Uniform distribution over an interval of up to 2 km/s at each depth. A lower velocity bound of 1.5 km/s is imposed to ensure the velocity is higher than the acoustic velocity in water.}
	\label{fig:true_model}
\end{figure}

We demonstrate the method by interrogating the size of a subsurface structure. Such questions appear frequently, such as where we wish to estimate the size of a subsurface ore body, or of a reservoir for carbon capture and storage, or to estimate the size of a volcanic magma chamber. In order to answer this question we choose to use seismic full-waveform inversion (FWI) to estimate the subsurface seismic velocity structure, and we infer the size of a subsurface structure defined by velocity anomalies. We use a part of the Marmousi model \citep{martin2006marmousi2} as the true velocity structure to demonstrate the method, and simulate 10 sources at 20 m depth in the water layer with 200 equally spaced receivers at a depth of 360 m across the horizontal extent of the seabed (Figure \ref{fig:true_model}a). This acquisition geometry may not be the optimal design, but it represents geometries that occur in reality where we have fixed seismometers and wish to answer specific scientific questions about the Earth's interior.

The model is discretised in space using a regular 200 $\times$ 120 grid of constant-velocity cells; the set of cell velocities constitutes parameters $\bm \uptheta$. The model's relationship to data is a full waveform simulation from each source to all receivers through velocity structure $\bm \uptheta$. The prior pdf of the velocity is assumed to be a Uniform distribution at each depth (Figure \ref{fig:true_model}b). We generate two waveform datasets using Ricker wavelets with dominant frequency of 4 Hz and 10 Hz respectively, so that the information gained from low frequency data and high frequency data can be compared. To obtain the posterior distribution of the velocity, we use SVGD to solve the inverse problem. SVGD is a variational inference method which updates a set of samples of parameter space (called particles) to minimize the difference between the pdf represented by their distribution and the true posterior pdf \citep{liu2016stein, zhang2020variational}. Details of the inversion procedure can be found in \citet{zhang2021bayesian}. Here we focus on the dark-blue triangular structure around X = 3.2 km and Z=1.0 km (red box in Figure \ref{fig:true_model}a). 

\begin{figure}
	\includegraphics[width=.9\linewidth]{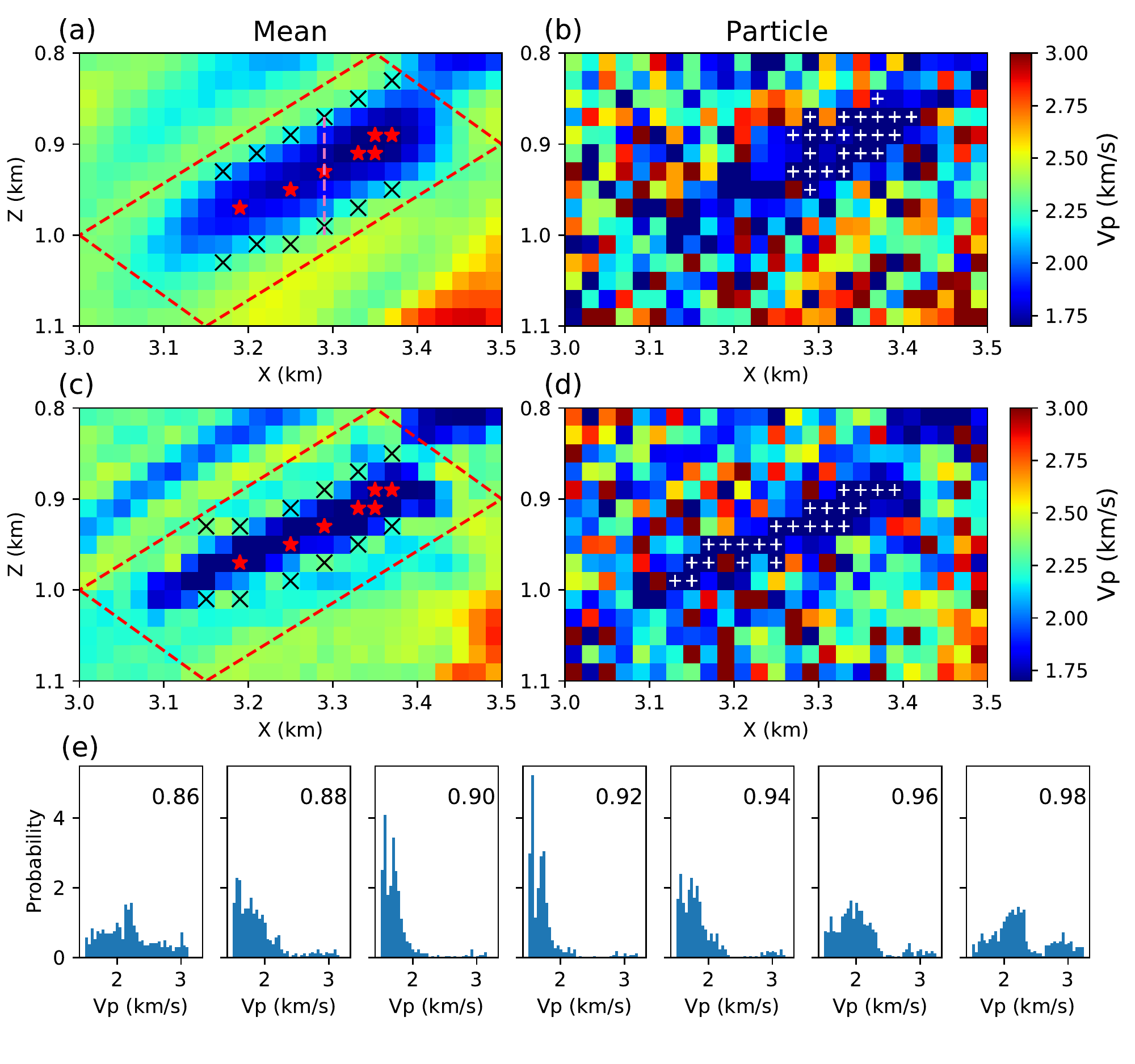}
	\caption{\textbf{(a)} Mean and \textbf{(b)} a random posterior particle (sample) obtained using the low frequency data. \textbf{(c)} Mean and \textbf{(d)} a random posterior particle obtained using the high frequency data. Red stars and black crosses denote locations that most likely have low velocity and high velocity respectively. The red dashed box shows the region where interrogation is performed. White pluses in (b) and (d) show continuous low velocity anomalies found for each particle. \textbf{(e)} Marginal distributions at X=3.29 km and at regular intervals across the depth range from 0.86 km to 0.98 km (purple dashed line in (a)) obtained using the low frequency data.}
	\label{fig:mean_particle}
\end{figure}

\begin{figure}
	\includegraphics[width=1.\linewidth]{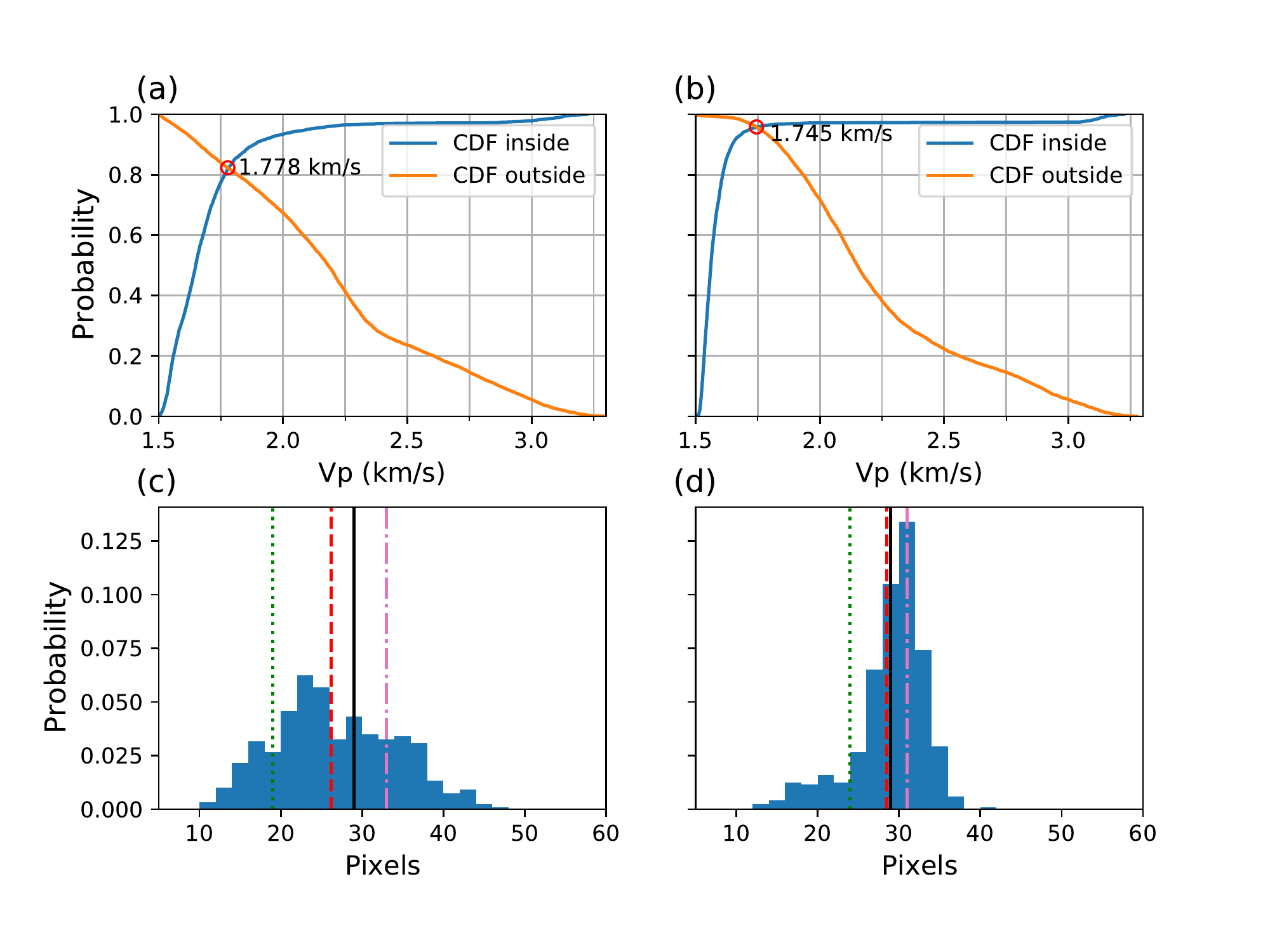}
	\caption{\textbf{(a)} The cumulative density functions (CDF) of points that are most likely within the low velocity anomaly accumulated in the positive Vp direction (blue line) and outside the low velocity anomaly accumulated in the negative Vp direction (orange line), both obtained using the low frequency data. \textbf{(b)} Same as (a) but obtained using the high frequency data. The red circle denotes the velocity value where the two CDFs have the same value, which is the threshold that discriminates low from high velocities with minimal bias. \textbf{(c)} and \textbf{(d)} show the distributions of the low velocity anomaly size obtained using the low and high frequency data, respectively. Red dashed line denotes the optimal answer obtained using interrogation theory (mean of the distribution -- equation \ref{eq:optimal_answer_simple}). Black line denotes the true size. For comparison, green dotted line and pink dash-dot line show the value obtained by directly interpreting the mean and median velocity structures, respectively.}
	\label{fig:cdf_answers}
\end{figure}

Figure \ref{fig:mean_particle}a and c show the posterior mean obtained using the low frequency data and high frequency data, respectively. In both results there is a low velocity anomaly of which we wish to estimate the size. In order to do this, we first define a "low velocity" by using a threshold: any cell whose velocity is smaller than the threshold is defined as a low velocity. Although the threshold may be chosen based only on the mean model, this procedure does not account for uncertainty in the velocity structure. For example, Figure \ref{fig:mean_particle}e shows the marginal distributions of velocity in cells at X=3.29 km across the vertical extent of the velocity anomaly (pink line in Figure \ref{fig:mean_particle}a). While the pixels at depth of 0.90 km and 0.92 km are highly likely to be within the low velocity anomaly, the pixels at depth of 0.86 km and 0.98 km clearly do not belong to this anomaly. In comparison, it is difficult to discriminate whether or not the pixels at depth of 0.88 km, 0.94 km and 0.96 km belong to the anomaly. This makes it difficult to choose an appropriate threshold with which the velocity anomaly can be defined. To resolve this issue, we select a set of points that are extremely likely to belong to the anomaly since their velocities are low in almost all particles (red stars in Figure \ref{fig:mean_particle}a and c) and another set of points that are highly likely to be outside of the anomaly (black crosses in Figure \ref{fig:mean_particle}a and c), each chosen using only the clearest results from cell marginal distributions. We then calculate the posterior cumulative density function (CDF) of the two sets of points, one accumulated in the positive and the other in the negative Vp direction, and plot them against each other. Figure \ref{fig:cdf_answers}a and b show the CDF plots obtained using the low frequency data and high frequency data, respectively. We define the velocity value where the two CDFs have the same probability as the threshold, because by definition of the CDF the probability that the velocity of those points within the anomaly are lower than this value equals the probability that the velocity of those points outside the anomaly are higher than this value. This velocity threshold therefore discriminates low from high velocities with minimal bias. For the low and high frequency examples in this study, the above procedure results in the thresholds 1.778 km/s and 1.745 km/s respectively.

%\begin{figure}
%	\includegraphics[width=1.\linewidth]{Figure_marginals.pdf}
%	\caption{Marginal distributions at X=3.29 km across the depth range from 0.86 km to 0.98 km (purple line in Figure \ref{fig:mean_particle}a) obtained using the low frequency data.}
%	\label{fig:marginals}
%\end{figure}

%\begin{figure}
%	\includegraphics[width=1.\linewidth]{Figure_answers.pdf}
%	\caption{The distribution of the low velocity anomaly size obtained using the \textbf{(a)} low and \textbf{(b)} high frequency data. Red dashed line denotes the optimal answer obtained using interrogation theory (mean of the distribution). Black line denotes the true size. For comparison, green dotted line and pink dash-dot line show the value obtained using the mean and median velocity structure respectively.}
%	\label{fig:answers}
%\end{figure}

We now define a low velocity anomaly as a continuous area whose velocity is smaller than the above threshold. Because of uncertainty in the posterior velocities (e.g. see Figure \ref{fig:mean_particle}b and d), there can be many such low velocity anomalies in a posterior velocity structure. To restrict ourselves to the main low velocity anomalies observed in the mean velocity structure, for each posterior particle we focus only on the largest continuous low velocity anomaly within the area of interest (red box in Figure \ref{fig:mean_particle}a and c). For example, Figure \ref{fig:mean_particle}b and d show examples of such low velocity anomalies which are denoted by white pluses. The size of each anomaly can then be computed, and in this study we simply use the number of interior pixels as the anomaly size. The above procedure of thresholding and counting pixels constitutes our target function $T(\bm \uptheta_m|m)$.

Figure \ref{fig:cdf_answers}c and d show the distributions of the target function $T(\bm \uptheta_m|m)$ (the anomaly size) obtained using the low frequency data and high frequency data, respectively. According to equation \ref{eq:optimal_answer_simple} the optimal answers are the mean of these distributions, which are denoted by dashed red lines in Figure \ref{fig:cdf_answers}c and d. For comparison the true size is denoted with black lines. The distribution of the anomaly size obtained using low frequency data is wider than that obtained using high frequency data, and the optimal answer obtained using only low frequency data also has a larger error. This demonstrates quantitatively and probabilistically that by using high frequency data we can obtain more accurate answers to specific scientific questions. In Figure \ref{fig:cdf_answers}c and d we also show the answers obtained by following the usual procedure of interpreting only the mean (green dotted line) and median (pink dash-dot line) velocity structures, using the same threshold value. The size obtained using the mean structure has the largest error, which clearly suggests that this structure provides less information about our question of interest. The size obtained using the median velocity structure has smaller error, probably because in this case the marginal pdfs are multimodal (Figure \ref{fig:mean_particle}e) and in such cases the median may represent the true structure better than the mean (which may lie between modes). Nevertheless, in both experiments, neither size is as accurate as those obtained using interrogation theory. And finally notice that none of these results could be obtained using only a single estimate of the velocity structure as then the minimal bias threshold cannot be properly defined. 

\section{Discussion}

The computational cost of constructing Bayesian solutions to scientific problems can be high. Particularly in imaging problems it is typical to communicate only statistics of the posterior pdf, often the mean or median, and some measure of uncertainty such as point-wise standard deviations. This study highlights the information loss incurred in such communications: not only are the answers derived above from the mean and median models relatively inaccurate, but they could not be calculated at all without additionally having samples of the posterior pdf in order to estimate an unbiased threshold that discriminates low velocity zones. We therefore advocate that methods to communicate the correlated structure of the posterior pdf are devised and used in future scientific communications. For example, one could publish not only statistics but also all samples of the posterior pdf that are computed as a matter of course by methods such as Monte Carlo and SVGD, or alternatively one could publish the parameters of solutions expressed as normalising flows \citep{zhao2020bayesian} or invertible neural networks \citep{zhang2021bayesianneuralnetwork} which then provide posterior samples almost for free. This would have the additional advantage that the same posterior pdf could be interrogated for answers to different questions thereafter. If this expectation of authors was widely adopted, the informational value of estimated posterior pdfs would increase, offsetting the cost of their computation.

In the above example we only used one model $m$, that is, seismic velocity structure and a forward wavefield simulator, and the distribution of parameter values was estimated using only one method of FWI. Equation \ref{eq:optimal_answer_simple} allows us to use multiple models. For example, different parameterizations can be included as different models in equation \ref{eq:optimal_answer_simple} to account for the uncertainty caused by specific parameterizations, and one can combine seismic velocity structures obtained using seismic methods with resistivity structures obtained using electromagnetic methods to answer the same questions. In addition, estimates of posterior pdfs obtained using different inference algorithms may also be combined to answer questions, such that the uncertainty caused by different algorithms can be taken into account in the procedure (see up-coming EIP report by Zhao et al., 2021). 

In this study we assumed a fixed experimental design which may not be an optimal design. To better quantify the answer to a scientific question, an optimal experimental design focussed on that question may also be found in the framework of interrogation theory \citep{arnold2018interrogation}. In reality after a round of interrogation the investigator may find that the answer to the question is not sufficiently constrained by the data. In such cases another interrogation can be conducted using an experimental design that is optimised based on the knowledge obtained in the first interrogation. This process can be repeated until a satisfactory answer is found.  

\section{Conclusions}

We used interrogation theory to quantify the size of a subsurface structure by interrogating the probabilistic results obtained from Bayesian seismic full-waveform inversion. The results demonstrated that the size obtained using interrogation theory provides an accurate estimate to the true size, which cannot be obtained using only one single velocity structure. This shows that the fully nonlinear uncertainty estimates are important for answering scientific questions, partly justifying their additional computational cost. We expect that the theory can be used to find answers for a range of real-world scientific questions, in particular for quantitative interpretation of geophysical inversion results to better understand the Earth.

\section*{Acknowledgments}
The authors thank the Edinburgh Imaging Project sponsors (BP, Schlumberger and Total) for supporting this research. This work has made use of the resources provided by the Edinburgh Compute and Data Facility (ECDF) (http://www.ecdf.ed.ac.uk/).

%\nocitep{*} 
\bibliographystyle{plainnat}
\bibliography{bibliography}

\appendix

\label{lastpage}

\end{document}